\documentclass[twocolumn,final,letterpaper,12pt,traditabstract]{swsc} 

\usepackage{natbib}
\usepackage{graphicx}
\usepackage{txfonts}
\usepackage{subfigure}
\usepackage{epstopdf}
\usepackage[backref]{hyperref}
\usepackage{lineno}

\hypersetup{colorlinks=true,citecolor=cyan,urlcolor=cyan,linkcolor=blue}

\def\rr#1{#1}
\def\rrr#1{#1}

\begin{document}


\title{Disturbances in the U.S. electric grid associated with geomagnetic activity}

\titlerunning{Grid disturbances associated with geomagnetic activity}

   \author{Carolus J. Schrijver
          \inst{1}
          \and
          Sarah D. Mitchell\inst{1}
          }

   \institute{ Lockheed Martin Advanced Technology Center, 
Palo Alto, CA,  USA\\
              \email{\href{mailto:schrijver@lmsal.com}{schrijver@lmsal.com}}
}

\abstract {Large solar explosions are responsible for space weather
  that can impact technological infrastructure on and around
  Earth. Here, we apply a retrospective cohort exposure analysis
    to quantify \rrr{the impacts of
  geomagnetic activity on the U.S. electric power grid for the period
  from 1992 through 2010. We find, with more than 3$\sigma$
  significance, that approximately 4\%\ of the disturbances in the
  U.S. power grid reported to the U.S.\ Department of Energy are
  attributable to strong geomagnetic activity and its associated
  geomagnetically induced currents.}}

   \keywords{solar magnetic activity -- geomagnetic disturbances --
     U.S. electric power grid -- geomagnetically induced currents}

   \maketitle


\section{Introduction}
Explosions powered by the Sun's magnetic field (``flares'' and
``coronal mass ejections'' or CMEs) are among the principal causes of
``space weather'' \citep[see, e.g.,][]{severeswx2008}.  These
electromagnetic storms can affect our technological infrastructure in
space, interfere with communications and GPS signals, and couple
through the geomagnetically induced currents (GICs) into the large-scale high-voltage
electric grid \citep[see, e.g.,][]{boteler+etal98,boteler+etal99,gaunt+coetzee2007}.
Despite the known impact of large space weather events on the
electrical power grid \citep[see,
e.g.,][]{severeswx2008,fema2010,kappenman2010,swximpactlloyds2011,jason2011}
-~including the 1989 Hydro-Qu{\'e}bec blackout
\citep{beland+small2004}~- relatively few studies of the general
correlation are available; case studies of individual events
\citep[such as by][]{kappenmanetal1997,kappenman2005} and compilations
of events for comparison with the solar cycle \citep[for example
by][]{boteler+etal98} generally focus on large storms and large
impacts.

There is a recognized hazard of catastrophic outages that may be
caused by geomagnetic superstorms larger than what we have experienced
in recent decades \citep{severeswx2008,fema2010,kappenman2010,hapgood2012}. Such
superstorms may cause trillions of dollars of damage
\citep{severeswx2008}, although it is acknowledged that such estimates
are rather uncertain \citep{jason2011}.  Other studies assessing the
economic impact on a statistical basis find significant correlations
between magnetometer data, GICs,
electric grid effects, and the conditions of the electric
  power grid market
\citep{forbesstcyr2004,forbesstcyr2008,forbesstcyr2010}. These
correlations are associated with market price variations on the order
of a few percent \citep{forbesstcyr2004}.

The main cause of GICs is the interaction of the geomagnetic field
with the magnetic field carried within CMEs and the surrounding
background magnetized solar wind that is modulated by them. With
speeds of $400 - 2500$\,km/s, it takes some $1-4$\,d for CMEs to
propagate from the Sun to the Earth, with a typical transit time of
$2-3$\,d. Correlations between the strength of CMEs, and the magnitude
of their impact in geospace continue to be studied, both
observationally and in numerical analyses
\citep[e.g.,][]{newelletal2007,schrijver2009b,2011JASTP..73..112A}. A
multitude of factors may play a role, including properties of the
solar events themselves and of the solar wind through which the events
travel to Earth
\citep[e.g.,][]{1973JGR....78...92R,2007LRSP....4....1P,schrijver+siscoe2010a}. Furthermore,
the magnitude of GICs depends on the location and time of day (through
the geomagnetic position relative to the Sun-Earth line) at impact, on
the structure of the magnetic field within the CME as \rr{that field interacts
with the magnetic field of the Earth during the CME's passage
(thereby inducing electric fields in the Earth} dependent on the direction and the
rate of change of the CME magnetic field), on the ground conductivities in
a wide area around any particular site for depths from ground level down
to in excess of 100\,km, and on the evolving architecture of the
electric power grid into which the induced electric field couples.

Within the electric power system, GICs can cause transformers to
operate in their nonlinear saturation range during half of the AC
cycle.  The consequences of half-cycle saturation include distortions
of the voltage pattern (reflected in the existence of harmonics to the
primary frequency), heating within the transformers, or
voltage-to-current phase shifts expressed as reactive power
consumption in the system \citep[see, e.g.,][ in these
proceedings]{laubyetal2013}.  The detection of harmonics or of strong
GICs may cause protective systems to trip, taking one or more
transformers off line to protect them from severe damage. The
implementation of such protective measures changes the grid's overall
configuration as well as the regional balance between power generation
and use which, in turn, can lead to power-quality variations in the
form of voltage and frequency swings \citep[causing, for example, the
1989 Hydro Qu{\'e}bec blackout, e.g.,][]{boteler2001}. Moreover, the
detection of GICs may cause system operators to change the operational
standards to project the overall system from damage, for example by
changing the transfer limits for power that may be transported between
segments of a grid (from where surplus power is more economically
available to regions where the demand is highest) to create a buffer
interval to keep GICs from pushing transformers into their nonlinear
range \citep[as is the standard ``GMD procedure'' during strong GIC
events for the PJM regional transmission operator on the east coast of
the US, see ][]{pjm101}.  Strong GICs can result in dissipative heating
within the transformers which may lead to their failure, either within
minutes or because of cumulative damage done over the life time of the
transformer \citep[e.g.,][in these proceedings]{gaunt2013}.

The strengths of GICs scale with the rate of change of the
geomagnetic field. As our study addresses the reliability of the
U.S. power grid, we chose to use a measure of geomagnetic variability
derived from geomagnetic measurements made around the central
latitudes of the U.S.  We verified that the use of a commonly-used
metric for large-scale geomagnetic variability, the Kp index, yields
the same results when allowing for the statistical uncertainties. We
find that even using criteria based directly on the occurrence of the
solar events that ultimately drive space weather yields the same
results, so that our findings are quite insensitive to the metric used
to quantify space weather conditions in which the \rrr{U.S. power grid
operates.}

\begin{figure*}
\noindent\includegraphics[width=\textwidth]{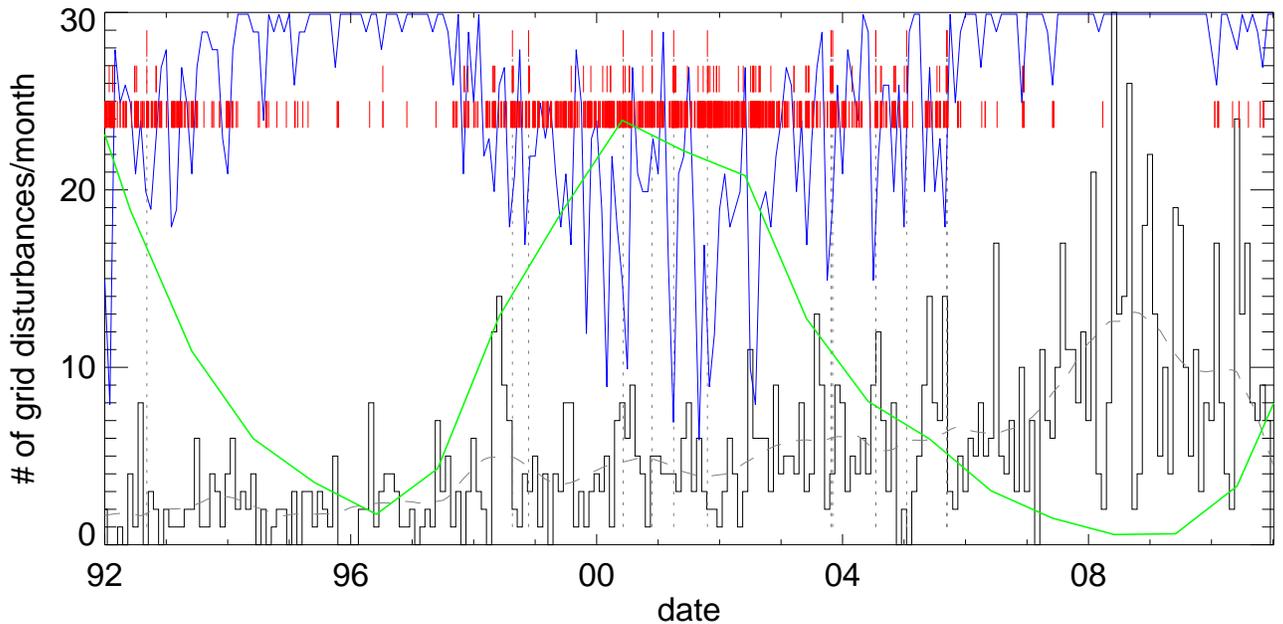}
\caption{\em Monthly frequency of grid disturbances (black histogram)
  and its 12-month running average (dashed gray line). Dates with (a)
  one of more M- or X-class flares, (b) one or more X-class flares, or
  (c) two or more X-class flares are marked by (red) bars near the top of the diagram. 
The monthly count of M and X class flares is shown inverted from the
top by the blue line only to avoid confusing overlaps between this and
the rate of grid disturbances, while still being able to view the
relative behavior in the peaks of each of the two curves.
The yearly sunspot number is overlaid as a (green) solid curve
(divided by a factor of five for display purposes).}
\label{fig:0}
\end{figure*}
\section{Disturbances in the U.S. power grid}\label{sec:griddata}
\rrr{As input to this study} we use a compilation
of ``system disturbances'' published annually by both the North
American Electric Reliability Corporation
(NERC\footnote{\hbox{http://www.nerc.com}}: available since 1992) and
by the Office of Electricity Delivery and Energy Reliability of the
Department of Energy
(DOE\footnote{\hbox{http://energy.gov/oe/office-electricity-delivery-and-}energy-reliability};
  available since 2000). 
NERC compiles this information for an
electric power market that serves over 300 million people throughout
the U.S.A.\ and in Ontario and New Brunswick in Canada, jointly
delivering power through more than 340,000\,km of high-voltage
transmission lines, linking 18,000 power plants within the U.S.\
\citep{jason2011}. 

The reported disturbances include, among others, ``electric service
interruptions, voltage reductions, acts of sabotage, unusual
occurrences that can affect the reliability of the bulk electric
systems, and fuel problems.'' The NERC reporting changed from
``selected disturbances'' to a more comprehensive listing starting in
2003 (following a grid collapse on August 14, 2003, affecting almost
50 million customers).  The DOE lists add information for 2008, 2010,
and 2011. To avoid a strongly inhomogeneous
data set, we exclude the DOE data for 2011 because of a marked change
in the types of events being reported on; for example, there are 79
events marked 'Vandalism' in 2011, which is $300\times$ the average
rate for that class of event reported in the 19 preceding years. We
thus use the combined disturbance
reports\footnote{http://www.nerc.com/page.php?cid=5$|$66}
\footnote{http://www.oe.netl.doe.gov/OE417$\_$annual$\_$summary.aspx}
for the 19-y period of 1992 through 2010.

We extracted the information on all 1216 disturbances listed in the
NERC-DOE reports, including the identified main cause, and the impact
on power and number of
customers affected (the latter two are often incompletely specified 
in the disturbance reports). Hence, our master list of attributed ``causes'' includes a
variety of weather conditions (storms, ice, lightning, etc.), operator
errors, equipment failures, transmission line faults, etc. 

Figure~\ref{fig:0} shows that the overall frequency of grid
disturbances exhibits a long-term increasing trend (the grey dashed
curve), modulated substantially on shorter time scales (shown on a
monthly basis by the black histogram). The figure also shows the
yearly-averaged sunspot number (green curve) that is -~as expected~-
clearly correlated with flare frequency (blue line; shown inverted
simply to avoid too much overlap with the grid-disturbance
frequency). No obvious correlation between solar flaring activity and
grid-disturbance frequency stands out (the peaks in the blue and black
curves do not align, nor do the dotted vertical lines -~dates of the
most severe solar activity with at least two X-class flares~- point to
particularly enhanced grid-disturbance frequencies), consistent with
our conclusions below that the effects are relatively weak, albeit
significant.

\section{Geomagnetic activity and electric power grid
  disturbances}\label{sec:geogrid}
As no direct attributions to space weather conditions have been made
for the events from the NERC-DOE reports studied here, we anticipate at most a weak effect by
space weather on the power grid that may be strongly modulated by
other processes affecting the grid's condition. Given enough
independent controlling variables (such as the evolving connectivities
within the power grid, the patterns of weather conditions, and the
grid loads and their changes with time around the country), one might
develop a multi-variate dependent variable model. However, insufficient
information is available to us at present: the detailed supply,
demand, and weather conditions are not included in the NERC/DOE
reports, and no information is available on the probability that no
reportable grid disturbances ensued from other operator errors, cases
of vandalism, or cyber attacks, for example. Moreover, as we find
below that only a few dozen disturbances in the sample of over 1,000
reported disturbances are attributable to enhanced space weather, we
cannot study separate grid areas while maintaining statistical
significance of the results. Such regional studies are natural
follow-up studies of this work, and those can focus not on the
statistical demonstration of susceptibility, as we do here, but on the
detailed physics of the electromagnetic coupling of GICs into the
power grid.

The power grid is generally operated in a state with enough power
being generated to meet customer demand, with only a relatively small
overcapacity --~the ``reserve margin''~-- available to accommodate for rapid changes in demand or to
compensate for ``contingency events'', i.e., external perturbations of the grid, such as lightning
strikes and other weather conditions, or internal events, such as component failures. Thus, whereas one might argue
that, for example, disturbances attributed to a lightning strike or to
an ice storm or to a heat wave might need to be removed from the list
of disturbances in a study looking to quantify the potential effects
of space weather, it may well be that the grid disturbance ensued only
because other factors, possibly including space weather, put the
system in a state of increased susceptibility. Taking this
perspective, the only disturbances that one might
exclude {\em a priori\/} are those that are attributed to planned
maintenance (provided these did not cause unforeseen disturbances
elsewhere) or to fuel shortages at the generating plants. Even cases
flagged as ``operator errors'' should not be excluded {\em a priori\/}
because the reports do not specify if the operators were responding to
changing grid conditions or merely to a truly local need to change the
operation of a grid segment. Even  ``vandalism'' might be
more or less effective in causing a grid disturbance depending on
system load and on the conditions of the geomagnetic field. In view of
the low numbers of events in the above sets, and as we do not wish to
inadvertently introduce biases in the process, we elected to work with
the complete set of reported grid disturbances.

\nocite{wacholderetal1992}
\nocite{morrisgardner1988}
\nocite{schulz+grimes2002}
\nocite{grimes+schulz2005}

\rr{Our study thus applies a standard method as used in, e.g.,
  epidemiology where it would be described as as a retrospective
  cohort exposure study with tightly matched controls (see, e.g.,
  Schulz and Grimes, 2002, on cohort vs. case-control studies). In our
  case, the cohort under study is the set of all dates from 1992
  through 2010. For all elements of that set, we study the "exposure"
  of the U.S. electric power grid to geomagnetic activity in excess of
  a specified threshold (in fact, three distinct thresholds based on
  percentiles of the distribution of geomagnetic-activity values, as
  defined below) and count the number of power-grid disturbances on
  such dates. The results of that are then compared to two control
  samples with distinctly different levels of ``exposure'', namely one
  with average exposure levels and another - the reference control
  sample - with low exposure levels.}

As the grid, its load, and its operating procedures change over time,
we need to devise a test that compares grid disturbance frequencies on
days of elevated geomagnetic activity to \rr{a control sample of} days of low geomagnetic
activity but with all other conditions being similar.  
\rr{We control against effects of continuously varying confounders
  that are associated with the evolution of the grid's infrastructure and
  operating rules over time by sample matching, specifically by
  ensuring time comparability of the "exposed" and control samples
  (e.g., Wacholder et al., 1992, and references therein): we form two
  control samples with matched frequencies by selecting dates near
  each of the dates of high "exposure" subject only to a criterion
  about their exposure levels. The selection of
  two control samples, rather than only one, provides additional insight into the effects at
  three different levels of exposure that can be compared within each
  selected exposure percentile, but we caution against comparison
  across percentiles because of the changes in grid operating
  conditions with time.}

\rr{In the definition of our control samples, we assume that
weather conditions, fuel prices, and vandalism, for example, are not
correlated with conditions on the Sun and in geospace within the 50-day sample matching windows (described below),} but that these
and other conditions form a background that varies independently of
solar and space weather. In view of the above, we adopt the following
avenue of research: we compare the frequency of grid disturbances
under severe space weather conditions with that under light
space-weather conditions, with the grid in otherwise similar
conditions. The second group is the control group containing grid
disturbances that are much less, if not entirely unaffected by space
weather. The contrast between these two samples enables us to estimate
\rr{the attributable risk, i.e., the impact of geomagnetic disturbances associated with} space weather.
 
\begin{table*}
\caption{Average daily frequency of grid disturbances for three
  distinct selection criteria: 
$g_a$ within the day of high geomagnetic activity  as measured by 
$|dB/dt|(30m)$;
$g_i$ for a day ending a 3-day period with the lowest average $|dB/dt(30m)|$ within 25 days of a day with high $|dB/dt(30m)|$;
$g_r$ for a day selected at random between 5 and 50 days before or after high $|dB/dt(30m)|$. The conditional criterion for days with high $|dB/dt(30m)|$ is
defined in the first column for each of the three rows.
The final column shows the total number of dates, $N_d$ with high
$|dB/dt(30m)|$ corresponding to the 2, 5, and 10 percentile levels. Uncertainties and subsampling criteria are as defined in Table~\ref{tab:1}.}\label{tab:3}
\begin{tabular}{l|c||c|c|r}
\hline
Selection criterion  for & $g_a$ [enhanced &  $g_i$ [low &  $g_r$ [nearby &  $N_d$ \\ 
reference dates & geomagn. act.] & nearby geomagn. act.] & random
date] & \\
\hline
\multicolumn{5}{c}{All disturbances (1216 cases)}\\
\hline
$|dB/dt(30m)| \ge 36.1$   & 0.230  $\pm$ 0.041 & 0.058  $\pm$ 0.020 & 0.136 $\pm$ 0.031 &   139 \\
$|dB/dt(30m)| \ge 24.5$   & 0.184  $\pm$ 0.023 & 0.107  $\pm$ 0.018 & 0.143 $\pm$ 0.023 &  347  \\
$|dB/dt(30m)| \ge 18.5$   & 0.167  $\pm$ 0.016 & 0.089  $\pm$ 0.012 & 0.147 $\pm$ 0.016 & 694 \\
\hline
\multicolumn{5}{c}{WET: Attributed to weather/external/technical causes
  (743 cases)}\\
\hline
$|dB/dt(30m)| \ge 36.1$ & 0.137 $\pm$ 0.031 & 0.043 $\pm$ 0.018 & 0.070 $\pm$ 0.024 &  \\
$|dB/dt(30m)| \ge 24.5$ & 0.115 $\pm$ 0.018 & 0.055 $\pm$ 0.013 & 0.077 $\pm$ 0.016 &  \\
$|dB/dt(30m)| \ge 18.5$ & 0.099 $\pm$ 0.011 & 0.050 $\pm$ 0.009 & 0.080 $\pm$ 0.010 &  \\
\hline
\multicolumn{5}{c}{U: Unclear/unknown attribution (473 cases)}\\
\hline
$|dB/dt(30m)| \ge 36.1$ & 0.094 $\pm$ 0.026 & 0.014 $\pm$ 0.010 & 0.066 $\pm$ 0.025 &  \\
$|dB/dt(30m)| \ge 24.5$ & 0.069 $\pm$ 0.010 & 0.052 $\pm$ 0.012 & 0.066 $\pm$ 0.017 &  \\
$|dB/dt(30m)| \ge 18.5$ & 0.068 $\pm$ 0.010 & 0.039 $\pm$ 0.007 & 0.068 $\pm$ 0.013 &  \\
\hline
\end{tabular}
\end{table*}

\begin{table*}
\caption{Average daily frequency of grid disturbances for three
  distinct selection criteria: 
$f_a$ from 2 to 5 days after a major flare;
$f_i$ for inactive intervals, i.e., 4-d intervals following the first
7-d intervals of no M or X flaring prior to  dates with major
flaring;
$f_r$ for a randomly-selected 4-d interval between 5 and 50 days before or after 'major
flaring'. The conditional criterion for days with 'major flaring' is
defined in the first column for each of the three rows.
The final two columns show the total number of dates, $N_d$,
and the total number of flares on such dates, $N_f$. Uncertainties
in $f_a$ and $f_i$  assume Poisson statistics; for $f_r$ 
the standard deviation of a sample of 100 random realizations is given.
Data are shown for all grid disturbances (top), for grid disturbances
attributed to weather, technical or external causes (center), and for
the complementary set of grid disturbances of unclear attribution (bottom). }\label{tab:1}
\begin{tabular}{l|c||c|c|r|r}
\hline
Selection criterion for& $f_a$ [2-5d after &  $f_i$ [nearby interval, &  $f_r$ [random &  $N_d$ &  $N_f$ \\ 
reference dates & M/X  flaring] & without M/X flaring] & nearby date] &
 & \\
\hline
\multicolumn{6}{c}{All disturbances (1216 cases)}\\
\hline
Multiple X flares             & 0.328  $\pm$ 0.072 & 0.063  $\pm$
0.031 & 0.210 $\pm$ 0.071 &   16 & 36 \\
At least one X flare         & 0.179  $\pm$ 0.020 & 0.116  $\pm$ 0.015
& 0.154 $\pm$ 0.022 &  116 & 136 \\
At least one M or X flare & 0.151  $\pm$ 0.006 & 0.126  $\pm$ 0.005 
& 0.148 $\pm$ 0.007 & 1054 & 1897\\
\hline
\multicolumn{6}{c}{WET: Attributed to weather/external/technical causes
  (743 cases)}\\
\hline
Multiple X flares         &  0.140 $\pm$ 0.047 &  0.031 $\pm$ 0.022 &  0.120 $\pm$ 0.056 & & \\
At least one X flare      &  0.071 $\pm$ 0.012 &  0.050 $\pm$ 0.010 &  0.085 $\pm$ 0.017 & & \\
At least one M or X flare &  0.077 $\pm$ 0.004 &  0.068 $\pm$ 0.004 &  0.083$\pm$ 0.005& & \\
\hline
\multicolumn{6}{c}{U: Unclear/unknown attribution (473 cases)}\\
\hline
Multiple X flares         &  0.188 $\pm$ 0.054 &  0.031 $\pm$ 0.022 &  0.090 $\pm$ 0.040 & & \\
At least one X flare      &  0.108 $\pm$ 0.015 &  0.067 $\pm$ 0.012 &  0.067 $\pm$ 0.011 & & \\
At least one M or X flare &  0.074 $\pm$ 0.004 &  0.058 $\pm$ 0.004 &  0.064 $\pm$ 0.004 & & \\
\hline
\end{tabular}
\end{table*}

\begin{figure*}
\noindent\includegraphics[width=\textwidth]{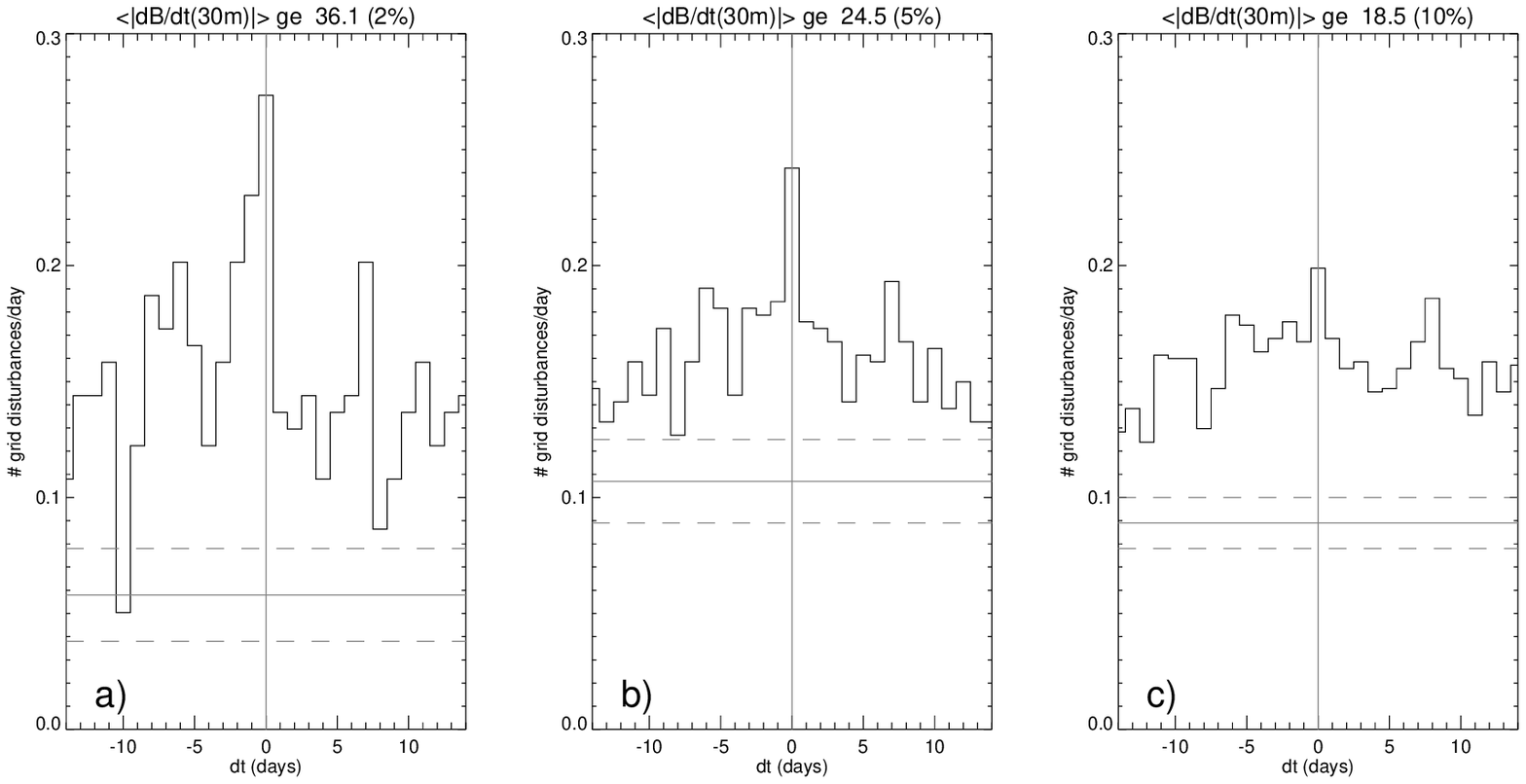}
\caption{\em Superposed epoch statistics of U.S.\ power grid
  disturbances for days of geomagnetic activity as measured by the
  maximum $dB/dt$ in 30-min.\ intervals, averaged for the U.S.\ BOU
  and FRD stations ($\langle |dB/dt(30m)| \rangle$) for the top 2, 5,
  and 10 percent of dates between 1992 and 2010, respectively. In each
  panel, the horizontal grey line and the dashed lines adjacent to it
  show the disturbance rates under quiescent space-weather conditions
  (i.e., the values of $g_i$, see Sect.~\ref{sec:geogrid} and Table~1)
  and the associated standard deviations, respectively. }
\label{fig:1}
\end{figure*}
To characterize the geomagnetic activity that may couple into the
U.S. power grid, we use data from the Boulder (BOU) and Fredericksburg
(FRD) stations\footnote{http://ottawa.intermagnet.org/apps/dl$\_$data$\_$def$\_$e.php}. With the minute-by-minute data in hand, we
compute the maximum value of $|dB/dt|$ for 30-min.\ intervals, for the
average of the two stations that are located along the central
latitudinal axis of the U.S., somewhat emphasizing the eastern U.S.\ as
do the grid and population that uses that.

In Table~1, we list the average grid disturbance rates, $g_a$, for
dates corresponding to the top $p=2, 5$, and $10$ percentiles of
geomagnetic activity, respectively. These numbers need to be compared
to disturbance rates in the absence of strong geomagnetic activity. In
order to ensure that the grid and its load are in a statistically
comparable state, we look at conditions in 50-d windows centered on
dates with high $|dB/dt(30m)|$. Selecting a random date within these
windows, but more than 5\,d away from the reference dates, yields the
disturbance rates $g_r$. These are lower than the rates $g_a$ for days
of high geomagnetic activity, but this selection criterion does not,
of course, avoid dates of significant geomagnetic activity. Hence, for
a second \rr{control} sample we select dates for the last day of the 3-d
interval of the lowest average $|dB/dt(30m)|$ within each of the 50-d
intervals. This yields disturbance rates $g_i$ for geomagnetically
inactive days. 

We note that for each of the percentile ($p$) levels, we find $g_a > g_r > g_i$, i.e., the disturbance frequency is highest within geomagnetically active days, lower for a randomly sampled nearby day, and lowest when geomagnetic activity is lowest. We caution that the values of $g_{a,r,i}$ for different $p$ levels are not directly comparable, because the coverage throughout the full sample period for each of these sets is different, and thus sensitive to long-term trends in grid, weather, and solar cycle. 

For each value of the
percentiles, $p$, we can estimate the number of grid disturbances in
excess of those occurring in conditions of low geomagnetic activity by
computing $N_p=(g_a-g_i) (p/100) n_d$ (where $n_d$ is the number of
days in our 19-y study interval): $N_p=24\pm 6, \, 27\pm 10, \, 54\pm
13$, respectively, for $p=2, 5, 10$. For higher $p$ values, more
disturbances may be associated with geomagnetic activity, but the
uncertainties on the values of $N_d$ rapidly increase (for $p=25$, for
example, the uncertainty in $N_{25}$ embraces $N_{10}$ within one
standard deviation), so that with the present data, we leave it at our
finding that at least $N_{10}\approx 50$ disturbances are attributable to
enhanced geomagnetic activity during the period of our study.

In order to assess whether our choice of metric for geomagnetic
variability would significantly bias the results, we repeated our
analysis for another commonly used index to characterize the
interaction of the geomagnetic field with the variable solar wind,
namely the Kp index. Kp is measured in sub-auroral mid-latitude
stations characteristic of activity in central regions of Europe and
the northern US, which is to be contrasted to the higher latitudes
used for the AE index or the more global distribution of stations used
for the Dst index.  The Kp index is determined from the variability of
the Earth's magnetic field, as measured by a network of ground-based
magnetometers, on a 3-h basis, expressed relative to quiet-day
variability on a scale from 0 to 9.  Analyzing daily averages of Kp,
we find results that are statistically consistent with
those in Table~1 based on $|dB/dt(30m)|$; we omit that table here for brevity.

As a final test, we
compare the compiled data base on disturbances in the electric power
grid to the catalog of solar flares maintained by NOAA, selecting only
large flares of GOES classes M and X (based on the logarithmic
$1-8$\AA\ peak brightness, such that an X1 flare is ten times brighter
than an M1 flare, and close to ten times more energetic overall
\citep{2002A&A...382.1070V}). For the period 1992-2010 there were 1897
M- and X-class flares on 1054 distinct dates. Nearly half of all
M-class flares and over 90\%\ of X-class flares are associated with
CMEs \citep[see the review by][and references therein]{schrijver2009b}
and thus most such flares affect the dynamics of the heliospheric
field, and thereby can couple into the geomagnetic field if directed towards the Earth.

We determined the grid disturbance frequencies $f_{a,i,r}$ using
three distinct selection criteria: (1) $f_a$ for intervals 2-5\,d
after major solar flaring (allowing for a range of CME propagation
times and a $1-2$\,d period of ensuing geomagnetic activity as the CME
passes Earth), (2) $f_r$ for 4-d intervals 
randomly selected within 50\,d of major solar flaring (in order to
remain reasonably within similar conditions for the grid otherwise)
but not within 5\,d of that flaring,
and (3) $f_i$ for the first 4-d intervals prior to the selected
reference dates of major solar flaring that 
end 7-d intervals of no major solar flaring, thus
selecting periods of relatively quiescent conditions in heliosphere and geospace. When
selecting dates for all X- or M-class flares, Table~\ref{tab:1} shows
$f_a=0.151\pm 0.006$\,disturbances/day and $f_i=0.126\pm 0.005$ (with the uncertainties
based on the numbers of events and assuming Poisson statistics).
We thus find a substantial increase in the frequency of grid disturbances
in the days following major flaring relative to quiescent intervals,
at a significance of about $4.5\sigma$.

Note that $f_r$ is not significantly different from $f_a$: with 1054
days of X or M flaring mostly concentrated around cycle maximum (see
Fig.~\ref{fig:0}), randomly selecting a date within 50\,d from a flare
frequently results in a date only days after another such major
flare. When we select only days with at least one X-class flare, the
chance of such overlaps is lowered: we see that in this case $f_a$ exceeds $f_r$ by
about $2\sigma$, while $f_a$ exceeds $f_i$ by $3.6\sigma$. Dates with
more than one X flare show an even more pronounced difference, but the
sample is relatively small and the uncertainties correspondingly
larger.

In order to estimate the total number $N_\odot$ of grid disturbances
added to the background grid variability by solar activity, we
multiply $f_a-f_i$ by the number of independent dates found within the
set of 4-d periods 2\,d after major flaring, yielding $N_\odot=50\pm
16$, or $4.1\pm 1.3$\%\ of all disturbances.

\rr{The study methodology applied above enforces a strict exclusion of information bias in creating the
  sample and its controls by ignoring the stated reason for a power
  grid disturbance in the reports. This is effective in eliminating
  confounders related to the reporting completeness and accuracy, and
  allows us to quantify the impact of a single variable among all
  possible impacts on the US power grid, namely the grid's "exposure"
  to geomagnetic activity (see, e.g., Grimes and Schulz, 2005, on
  selection biases in samples and their controls, specifically their
  example on pp. 1429-1430). It is instructive, however, to see the
  impact of introducing a selection bias by a coarse separation of
  identified causes.}

Tables~\ref{tab:3} and~\ref{tab:1}  \rr{show the grid disturbance
frequencies when separating the disturbances} into two broad categories.  One
category (WET) contains clear attributions to weather (including hot
and cold weather, wind, ice, and lightning; 637 entries), external
factors (fires, sabotage, earthquakes, collisions, etc.; 63), and
technical issues (fuel shortages, maintenance, etc.; 43 entries). The
complementary list (U, with 473 entries) shows causes such as 'line
fault', 'operator error', 'public appeal', 'voltage reduction', 'load
shed', 'equipment failure', etc., for which no clear correlation with
weather, external, or technical issues is listed.  The contrasts
between $g_a$ and $g_i$ for days with geomagnetic activity in the top
percentiles for the both types of events are statistically comparable
to those of the full sample. The same is true for the contrast between
the conditional grid-disturbance frequencies given flare activity,
i.e., $f_a/f_i$ for high, medium, and moderate flaring
activity. \rr{We conclude from this experiment that the susceptibility
  of the US power grid appears to be statistically similar to
  geomagnetic activity for the two classes of causes, and that our
  findings would thus have been identical had we focused only on those
  disturbances for which the identified cause is clearly proximate (as
  follows from the examples of included causes in group U given above).}

The results in Tables~\ref{tab:3} and~\ref{tab:1}   are a direct demonstration of the
statistically significant impact of geomagnetic activity on the US
power grid. We add to that the simple visualization in
Figure~\ref{fig:1} which emphasizes this impact in a slightly
different manner.  As the space-weather effects on the U.S.\ power
grid over our 19-y interval are relatively weak, we use a superposed
epoch analysis to visualize the magnitude of the
effects. Figure~\ref{fig:1} shows the average grid disturbance
frequencies for days with geomagnetic activity, as measured by
$|dB/dt(30m)|$, in the top 2, 5, and 10 percentiles in panels {\em a},
{\em b}, and {\em c}, respectively, for 4-week periods centered on
those most active dates. There clearly is a peak on the central dates
relative to their surrounding periods, revealing a dependence of the
U.S.\ power grid reliability on space-weather conditions. Often, solar
active regions exhibit series of flaring and coronal mass ejections
over periods of multiple days, sometimes up to a full two weeks as a
flare-productive region crosses the disk. Hence, for a comparison of
the geomagnetically active dates with a reference date of low
geomagnetic activity, the curves shown in Fig.~\ref{fig:1} do not
provide suitable information to set a baseline level for grid
disturbances in periods of low geomagnetic activity; that baseline
level was discussed above and presented in the Tables.

\begin{figure}
\noindent\includegraphics[width=9cm]{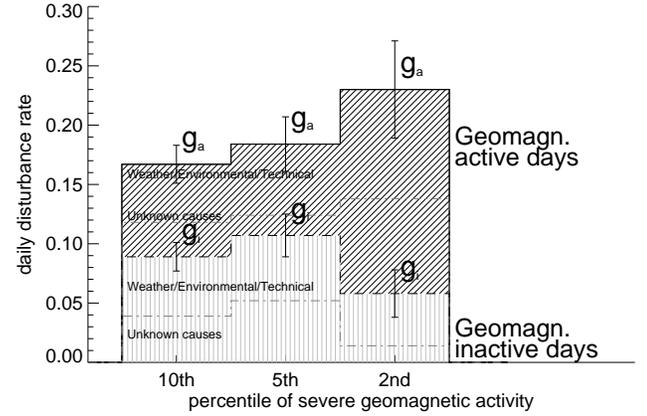}
\caption{\em Graphical rendition of the grid disturbance rates under
  different levels of geomagnetic activity, as listed in Table~1. The
  three columns show the results for geomagnetic activity in the 10th,
5th, and 2nd upper percentiles of geomagnetic activity as measured by
$|dB/dt(30m)|$, respectively. The excess in event frequency on geomagnetically
active days compared to nearby inactive days is shown by dark diagonal shading. }
\label{fig:2}
\end{figure}
In conclusion, we find a statistically significant enhancement in the
frequency of power grid disturbances on days of high geomagnetic
activity, regardless of which measure for geomagnetic activity we use:
a metric for 30-min.\ variability characteristic of the central
U.S.A. (for which the results from Table~1 are shown graphically in
Fig.~\ref{fig:2}), a metric for the 3-h (Kp) variability for high
latitudes around the globe, or when looking at intervals following
days of major solar flaring. This enhancement means that at least
$\approx 4$\%\ of reported grid disturbances are attributable in whole
or in part to enhanced geomagnetic activity. We note that although
significant, the fraction of grid disturbances that we find
attributable to GIC effects is relatively small, so that the overall
number of disturbances attributable to space weather is small even
during periods of severe solar activity: even on days with the most
extreme geomagnetic activity, only $\approx ((0.23\pm 0.04)-(0.06\pm
0.02))=0.17\pm 0.05$ disturbances per day would expected in
association with severe space weather \rrr{(using numbers from
Table~\ref{tab:3}).}

\section{Discussion}
\rr{\rrr{We perform a retrospective cohort study} to quantify the
  susceptibility of the US power grid to disturbances attributable (at
  least in part) to geomagnetic activity. The results of such a study can be expressed
  in a simple contingency table, of which Tables~\ref{tab:3} and~\ref{tab:1} are
  variants. Alternatively, it could be expressed as odds ratios and
  their confidence intervals (e.g., Morris and Gardner, 1988) or in
  terms of chi square, both of which can be derived from the numbers
  in Tables~\ref{tab:3} and~\ref{tab:1}. Such numbers convey the same message as the
  combined contingency Tables 1 and 2: the impact of the "exposure" is
  statistically significant to more than the 3-sigma level, i.e., the
  null hypothesis that the US power grid is insensitive to space
  weather is rejected with more than 0.975 (or 32 in 33) probability. }

Except in rare cases, solar energetic events and resulting geomagnetic
activity are not presently recognized as contributing to power grid
disturbances. In fact, no grid disturbance was thus attributed over
the 19-y period studied, either as primary cause or as contributing
factor, in the NERC-DOE reports.  This is to be contrasted to our
finding (significant in excess of 4 standard
deviations) that over the 19-y period of our study, $\approx 50$ grid
disturbances reported to NERC and DOE had strong geomagnetic and solar
activity as a contributing factor.

The present lack of recognition of geomagnetic activity as a
contributing agent in grid disturbances may reflect that, in contrast
to extreme storms, moderate to severe space weather conditions do not
by themselves cause such disturbances but instead \rrr{are one factor
among all others to which the electric power grid is susceptable.}
These other perturbations may be identified as the
cause of the disturbance, but our study leads us to conclude that
sometimes geomagnetic activity is a contributing factor. One may think
of parallels such as the activity of skiers that contributes to the
triggering of avalanches particularly if conditions of snowfall and
weather are right; or one may consider the effect of being engaged in
cell phone calls on the likelihood of vehicular accidents in demanding
traffic conditions. We conjecture that in the grid disturbances that
we find to be influenced by geomagnetic activity \rrr{and their induced
currents,} this activity may be
the equivalent of the presence of a skier or of being on the phone in
the above analogies.  The U.S.\ power grid is, after all, a highly
complex coupled system in which initially localized problems can
cascade into disturbances of any size (characterized on the large end
of the spectrum by a scale-free power-law distribution typical of
nonlinear systems \citep{carreras+etal2003,talukdar+etal2003}),
compounded by the fact that GICs induced by space weather extend over
a large fraction of the footprint of the U.S.\ electric power grid and
thus can have effects in various locations simultaneously.

\rrr{The apparent correlation of electric power grid disturbances}
with pronounced solar and geomagnetic activity warrants the
investigation and implementation of mitigation strategies and the
support of a space weather research program as well the continued
development of a space weather forecasting system.  Such an investment
would also help us to better understand what protection society would
need if faced with more severe space weather than experienced in
recent decades, or from more extensive cascading effects in our
ever-more coupled technological infrastructure.


\begin{acknowledgements}
  This work was supported by Lockheed Martin Independent Research
  funds.  We thank D.\ Boteler, D.\ Chenette, K.\ Forbes, M.\ Hapgood,
  L.\ Lanzerotti, R.\ Lordan, and A.\ Title for discussions,
  and them and the reviewers for helpful suggestions. The results
  presented in this paper rely on data collected at magnetic
  observatories. We thank the national institutes that support them
  and INTERMAGNET for promoting high standards of magnetic observatory
  practice (www.intermagnet.org).
\end{acknowledgements}


\begin{thebibliography}{32}
\expandafter\ifx\csname natexlab\endcsname\relax\def\natexlab#1{#1}\fi

\bibitem[{{\it {Andreeova} et~al.\/}(2011){\it {Andreeova}, {Pulkkinen},
  {Palmroth}, and {McPherron}\/}}]{2011JASTP..73..112A}
{Andreeova}, K., T.~I. {Pulkkinen}, M.~{Palmroth}, and R.~{McPherron},
  {Geoefficiency of solar wind discontinuities}, {\it Journal of Atmospheric
  and Solar-Terrestrial Physics\/}, {\it 73\/}, 112--122, 2011,
  doi:10.1016/j.jastp.2010.03.006.

\bibitem[{{\it B{\'e}land and Small\/}(2004)}]{beland+small2004}
B{\'e}land, J., and K.~Small, {Space weather effects on power transmission
  systems: the cases of Hydro-Qu{\'e}bec and Transpower New Zealand Ltd}, in
  {\it Effects of space weather on technological infrastructure\/}, edited by
  I.~A. Daglis, pp. 287--299, Kluwer Academic Publishers, The Netherlands,
  2004.

\bibitem[{{\it Boteler\/}(2001)}]{boteler2001}
Boteler, D.~H., {Space weather effects on power systems}, in {\it Space
  Weather, Geophys. Monogr. Ser., Vol. 125\/}, edited by P.~Song, H.~J. Singer,
  and G.~L. Siscoe, pp. 347--352, AGU, Washington, D.C., 2001.

\bibitem[{{\it Boteler and {Jansen van Beek}\/}(1999)}]{boteler+etal99}
Boteler, D.~H., and G.~{Jansen van Beek}, {August 4, 1972 revisited: a new look
  at the geomagnetic disturbance that caused the L4 cable system outage}, {\it
  Geophys. Res. Lett.\/}, {\it 26\/}, 577--580, 1999.

\bibitem[{{\it Boteler et~al.\/}(1998){\it Boteler, Pirjola, and
  Nevanlinna\/}}]{boteler+etal98}
Boteler, D.~H., R.~J. Pirjola, and H.~Nevanlinna, {The effects of geomagnetic
  disturbances on electrical systems at Earth's surface}, {\it Adv. Space
  Res.\/}, {\it 22\/}, 17--27, 1998.

\bibitem[{{\it Carreras et~al.\/}(2003){\it Carreras, Lynch, Newman, and
  dobson\/}}]{carreras+etal2003}
Carreras, B., V.~Lynch, D.~Newman, and I.~dobson, {Blackout mitigation
  assessment in power transmission lines}, {\it System Sciences\/}, 2003,
  doi:10.1109/HICSS.2003.1173911, doi10.1109/HICSS.2003.1173911.

\bibitem[{{\it FEMA\/}(2010)}]{fema2010}
FEMA, M., NOAA, {\it {Managing critical disasters in the transatlantic domain -
  the case of a geomagnetic storm}\/}, FEMA, Washington, DC, 2010.

\bibitem[{{\it Forbes and {St. Cyr}\/}(2004)}]{forbesstcyr2004}
Forbes, K.~F., and O.~C. {St. Cyr}, {Space weather and the electricity market:
  an initial assessment}, {\it Space Weather\/}, {\it 2\/}, S10,003, 2004,
  doi:10.1029/2003SW000005.

\bibitem[{{\it Forbes and {St. Cyr}\/}(2008)}]{forbesstcyr2008}
Forbes, K.~F., and O.~C. {St. Cyr}, {Solar activity and economic fundamentals:
  evidence from 12 geographically disparate power grids}, {\it Space
  Weather\/}, {\it 6\/}, S10,003, 2008, doi:10.1029/2007SW000350.

\bibitem[{{\it Forbes and {St. Cyr}\/}(2010)}]{forbesstcyr2010}
Forbes, K.~F., and O.~C. {St. Cyr}, {An anatomy of space weather's electricity
  market impact: case of the PJM power grid and the performance of its 500 kV
  transformers}, {\it Space Weather\/}, {\it 8\/}, S09,004, 2010,
  doi:10.1029/2009SW000498.

\bibitem[{{\it Gaunt\/}(2013)}]{gaunt2013}
Gaunt, C.~T., {Reducing uncertainty - an electricity utility response to severe
  solar storms}, {\it Journal of Space Weather and Space Climate\/}, 2013,
  submitted.

\bibitem[{{\it Gaunt and Coetzee\/}(2007)}]{gaunt+coetzee2007}
Gaunt, C.~T., and G.~Coetzee, {Transformer failures in regions incorrectly
  considered to have low GIC-risk}, in {\it IEEE PowerTech\/}, IEEE, Lausanne,
  2007.

\bibitem[{{\it Grimes and Schulz\/}(2005)}]{grimes+schulz2005}
Grimes, D.~A., and K.~F. Schulz, {Compared to what? Finding controls for
  case-control studies}, {\it The Lancet\/}, {\it 365\/}, 1429--1433, 2005.

\bibitem[{{\it Hapgood\/}(2011)}]{swximpactlloyds2011}
Hapgood, M., {\it {Lloyd's 360$^\circ$ risk insight: Space Weather: Its impacts
  on Earth and the implications for business}\/}, Lloyd's, London, UK, 2011.

\bibitem[{{\it Hapgood\/}(2012)}]{hapgood2012}
Hapgood, M., {Prepare for the coming space weather storm}, {\it Nature\/}, {\it
  484\/}, 311--313, 2012, doi:10.1038/484311a.

\bibitem[{{\it JASON\/}(2011)}]{jason2011}
JASON, {\it {Impacts of severe space weather on the electric grid
  (JSR-11-320)}\/}, The MITRE Corporation, McLean, VA, 2011.

\bibitem[{{\it Kappenman\/}(2010)}]{kappenman2010}
Kappenman, J., {\it {Geomagnetic storms and their impacts on the US power grid
  (Tech. Rep. Meta-R-319}\/}, Metatech Corp., Goleta, CA, 2010.

\bibitem[{{\it Kappenman\/}(2005)}]{kappenman2005}
Kappenman, J.~G., {An overview of the impulsive geomagnetic field disturbances
  and power grid impacts associated with the violent Sun-Earth connection
  events of 29–31 October 2003 and a comparative evaluation with other
  contemporary storms}, {\it Space Weather\/}, {\it 3\/}, 2005,
  doi:10.1029/2004SW000128.

\bibitem[{{\it Kappenman et~al.\/}(1997){\it Kappenman, Zanetti, and
  Radasky\/}}]{kappenmanetal1997}
Kappenman, J.~G., L.~J. Zanetti, and W.~A. Radasky, {Geomagnetic storm
  forecasts and the power industry}, {\it Eos Trans. AGU\/}, {\it 78\/}, 37,
  1997.

\bibitem[{{\it Lauby et~al.\/}(2013){\it Lauby, Moura, and
  Rollinson\/}}]{laubyetal2013}
Lauby, M., J.~Moura, and E.~Rollinson, {Effects of geomagnetic disturbances on
  the north American bulk power system}, {\it Journal of Space Weather and
  Space Climate\/}, 2013, submitted.

\bibitem[{{\it Morris and Gardner\/}(1988)}]{morrisgardner1988}
Morris, J.~A., and M.~J. Gardner, {Calculating confidence intervals for
  relative risks (odds ratios) and standardized ratios and rates}, {\it British
  Medical Journal\/}, {\it 296\/}, 1313--1316, 1988.

\bibitem[{{\it Newell et~al.\/}(2007){\it Newell, Sotirelis, Liou, Meng, and
  Rich\/}}]{newelletal2007}
Newell, P., T.~Sotirelis, K.~Liou, C.~I. Meng, and F.~J. Rich, {A nearly
  universal solar wind-magnetosphere coupling function inferred from 10
  magnetospheric state variables}, {\it JGR\/}, {\it 112\/}, 2007,
  doi:10;1029/2006JA012015.

\bibitem[{{\it {PJM State and Member Training Dept.}\/}(2010)}]{pjm101}
{PJM State and Member Training Dept.}, {\it {Weather and Environmental
  Emergencies}\/}, PJM, : http://pjm.acrobat.com/p37769123, 2010.

\bibitem[{{\it {Pulkkinen}\/}(2007)}]{2007LRSP....4....1P}
{Pulkkinen}, T., {Space Weather: Terrestrial Perspective}, {\it Living Reviews
  in Solar Physics\/}, {\it 4\/}, 1, 2007.

\bibitem[{{\it {Russell} and {McPherron}\/}(1973)}]{1973JGR....78...92R}
{Russell}, C.~T., and R.~L. {McPherron}, {Semiannual variation of geomagnetic
  activity.}, {\it JGR\/}, {\it 78\/}, 92--108, 1973,
  doi:10.1029/JA078i001p00092.

\bibitem[{{\it Schrijver and Siscoe\/}(2010)}]{schrijver+siscoe2010a}
Schrijver, C., and G.~Siscoe, {\it {Heliophysics. Space storms and radiation:
  Causes and Effects}\/}, Cambridge University Press, Cambridge, U.K., 2010.

\bibitem[{{\it {Schrijver}\/}(2009)}]{schrijver2009b}
{Schrijver}, C.~J., {Driving major solar flares and eruptions: A review}, {\it
  Advances in Space Research\/}, {\it 43\/}, 739--755, 2009,
  doi:10.1016/j.asr.2008.11.004.

\bibitem[{{\it Schulz and Grimes\/}(2002)}]{schulz+grimes2002}
Schulz, K.~F., and D.~A. Grimes, {Case-control studies: research in reverse},
  {\it The Lancet\/}, {\it 359\/}, 431--434, 2002.

\bibitem[{{\it {Space Studies Board}\/}(2008)}]{severeswx2008}
{Space Studies Board}, {\it {Severe space weather events - understanding
  societal and economic impacts}\/}, National Academy Press, Washington, D.C.,
  U.S.A., 2008.

\bibitem[{{\it Talukdar et~al.\/}(2003){\it Talukdar, Apt, Ilic, Lave, and
  Morgan\/}}]{talukdar+etal2003}
Talukdar, S., J.~Apt, M.~Ilic, L.~Lave, and M.~Morgan, {Cascading failures:
  survival vs.\ prevention}, {\it The Electricity Journal\/}, {\it 16, Issue
  9\/}, 25--31, 2003.

\bibitem[{{\it {Veronig} et~al.\/}(2002){\it {Veronig}, {Temmer}, {Hanslmeier},
  {Otruba}, and {Messerotti}\/}}]{2002A&A...382.1070V}
{Veronig}, A., M.~{Temmer}, A.~{Hanslmeier}, W.~{Otruba}, and M.~{Messerotti},
  {Temporal aspects and frequency distributions of solar soft X-ray flares},
  {\it A{\&}A\/}, {\it 382\/}, 1070--1080, 2002,
  doi:10.1051/0004-6361:20011694.

\bibitem[{{\it Wacholder et~al.\/}(1992){\it Wacholder, Silverman, McLaughlin,
  and Mandel\/}}]{wacholderetal1992}
Wacholder, S., D.~T. Silverman, J.~K. McLaughlin, and J.~S. Mandel, {Selection
  controls in case-control studies}, {\it American Journal of Epidemiology\/},
  {\it 135\/}, 1042--1050, 1992.

\end{thebibliography}


\end{document}